# Energy management system for biological 3D printing by the refinement of manifold model morphing in flexible grasping space

Kang WANG [a,b,*]

[a] School of Nursing, The Hong Kong Polytechnical University, Kowloon, Hong Kong SAR

[b] School of Mechanical Engineering, Zhejiang University, Hangzhou, 310027, China

**Abstract:** The use of 3D printing, or additive manufacturing, has gained significant attention in recent years due to its potential for revolutionizing traditional manufacturing processes. One key challenge in 3D printing is managing energy consumption, as it directly impacts the cost, efficiency, and sustainability of the process. In this paper, we propose an energy management system that leverages the refinement of manifold model morphing in a flexible grasping space, to reduce costs for biological 3D printing. The manifold model is a mathematical representation of the 3D object to be printed, and the refinement process involves optimizing the morphing parameters of the manifold model to achieve desired printing outcomes. To enable flexibility in the grasping space, we incorporate data-driven approaches, such as machine learning and data augmentation techniques, to enhance the accuracy and robustness of the energy management system. Our proposed system addresses the challenges of limited sample data and complex morphologies of manifold models in layered additive manufacturing. Our method is more applicable for soft robotics and biomechanisms. We evaluate the performance of our system through extensive experiments and demonstrate its effectiveness in predicting and managing energy consumption in 3D printing processes. The results highlight the importance of refining manifold model morphing in the flexible grasping space for achieving energy-efficient 3D printing, contributing to the advancement of green and sustainable manufacturing practices.

---

[*]Corresponding author: Kang WANG (kangwang@polyu.edu.hk)

morphing; Flexible grasping space; Layered additive manufacturing.

# 1 Introduction

In recent years, three-dimensional printing (3DP) has attracted unprecedented attention due to its potential for revolutionizing traditional manufacturing processes [1,2]. Compared with traditional production methods, 3D printing has greatly improved the success rate of manufacturing, which is thus also called rapid prototyping [3–5]. When a CAD model is given, and additive manufacturing divides the model into several two-dimensional (2D) layers, producing these layers layer-upon-layer to make components with complicated structures. The layer-by-layer approach to manufacturing makes it possible to use additive manufacturing outside of the realm of product design [6,7]. With the help of smart material, 3D printing structure can transfer from one state to another under corresponding excitation, which is also termed 4D printing [8].

There is a growing body of literature that recognizes the importance of energy and consumption existing in additive manufacturing. Choy et al. [9] proposed that the distinctive deformation behavior of functionally graded materials (FGS) lattice structures was progressive and predictable, regardless of lattice design. Alomarah et al. [10] employed numerical models to study the effects of friction and energy absorption for re–entrant chiral auxetic structure. Habib et al. [11] proposed an in-plane energy absorption evaluation method of 3D printed polymeric honeycombs. Bonami et al. [12] have solved the mixed-integer nonlinear programming problem using a nonlinear programming-based branch-and-bound algorithm specifically tailored to the problem.

To optimize the fabrication energy efficiency and quality, a variety of researches have been undertaken for additive manufacturing from various perspectives [13–16]. For example, Diourté et al. [13] proposed a wire arc additive manufacturing strategy to minimize the start/stop phases of the arc to one unique cycle, by generating a continuous trajectory in spiral form for closed-loop thin parts. Allum et al. [14] proposed a non-planar geometries, called ZigZagZ, which deposits filaments simultaneously in the X, Y,

Z direction. Delfs et al. [17] developed a prediction method of the surface quality in dependence of the building orientation of a part, and the prediction results are thus be used to optimize the orientation to get a desired surface quality. To analyze the mechanical vibration in additive manufacturing system, Tlegenov et al. [18] proposed a nozzle condition monitoring technique in fused filament fabrication 3D printing using a vibration sensor. However, most of the existing methods are unable to fundamentally solve the vibration problem in additive manufacturing, since the vibration is an inherent attribute of mechanical process due to the intrinsic kinematic system, but only could be suppressed.

Recent studies have shown that toolpath planning becomes a promising method to provide optimal trajectory for kinematic system, suppressing the mechanical vibration in fundamental respects for additive manufacturing [19,20]. Dwivedi and Kovacevic [21] proposed an automated continuous toolpath planning approach, which connects individual zigzag paths in each decomposed infill sub-region. Griffiths [22] proposed a Hilbert's curve based toolpath strategy for machining curved surfaces. Kuipers et al. [23] proposed a adaptive width control strategy by using various beading schemes to reduce the bead width range. Etienne et al. [24] developed a curved slicing method to better align the object surface by generating a toolpath to adapt the variable slicing layer height. The mentioned-above researches contribute to the favorable development of additive manufacturing technologies.

The above published research promoted progress in additive manufacturing. However, most current methods are sensitive to the morphology complexity of the models which will reduce energy efficiency of 3DP system along with the increasing of facets amount and surface morphology. They lack an energy management system to evaluate their methods whether and how much reduce the energy consumption [25–27]. Currently, the success of deep neural networks in supervised learning tasks demonstrates they can be applied to optimizing Energy management system in 3DP. Note that deep neural networks not only depend on the depth of the network structure, but also rely on a large amount of annotated sample data and large-scale iterative training [28,29]. When the sample data for certain deep learning models is insufficient or the annotated samples are too few, underfitting or overfitting can easily occur.

It is observed that establishing a large and complete dataset for energy consumption using traditional data acquisition methods requires a significant amount of manpower and resources. In the field of computer vision, common data augmentation techniques such as shifting, rotating, scaling, cropping, and flipping are used to augment image datasets. However, conventional image data augmentation methods may result in the loss of intrinsic features, as the CAD models are 3D data and usually have entirely different distribution.

To address these issues, this paper proposes an energy management system for 3D printing by the refinement of manifold model morphing in flexible grasping space. The contributions are summarized as below:

- A manifold model morphing method in a flexible grasping space is proposed to generate multiple similar but distinct family-products. A grasping hand is adopted as an example, which mimics human hand grasping action.

- Kinematics model of self-grasping manifold is constructed to keep the manifold model morphing more like a robot hand grasping. It is able to realistically generate multiple models with different poses.

- These CAD models are sliced to generate augmented data, which is feed to a semi-supervised deep residual network to enhance the accuracy and robustness of the energy management system. Our method is more applicable for soft robotics and biomechanisms.

- We evaluate the performance of our system through extensive experiments and demonstrate its effectiveness in predicting and managing energy consumption in 3D printing processes. It demonstrates that our proposed system addresses the challenges of limited sample data and complex morphologies of manifold models in layered additive manufacturing.

## 2    Manifold model morphing in flexible grasping space

### 2.1 Grasping posture modeling using CT images

The surface machining precision and topography of a work piece is denoted by relevant parameters, such as the parameters of surface roughness, the parameters of waviness, the parameters of shape error and so on. Shape error and position error are referred to as geometric tolerance. The geometric tolerance can usually be obtained by macroscopic measurement of dimension error. Wave spacing λ refers to the distance between the adjacent two peaks or two troughs.

The actual topography is considered as the original signal, filtered by different wave spacing λ, surface roughness, waviness and shape error can be obtained by low pass filter H(λ), band-pass filter H(λ) and high-pass filter H(λ). Surface roughness is a component of surface texture. It is quantified by the deviations in the direction of the normal vector of a real surface from its ideal form. Roughness is typically considered to be the high-frequency, short-wavelength component of a measured surface. Surface waviness is the measurement of the more widely spaced component of surface texture. It is a broader view of roughness because it is more strictly defined as "the irregularities whose spacing λ is greater than that of the roughness". It can occur from machine or work deflections, chatter, residual stress, vibrations, or heat treatment. Waviness should also be distinguished from flatness, both by its shorter spacing and its characteristic of being typically periodic in nature. Waviness contains linear waviness and circumferential waviness. There are several parameters for expressing waviness height, the most common being Wa & Wt, for average waviness and total waviness, respectively. The geometric shape error can be divided into isolated error and associated error. The isolated error includes: flatness, roundness, cylindricity, line profile, planar profile. The associated error includes: parallelism, perpendicularity, inclination, coaxiality, symmetry, location, circular run-out, total run-out.

$$\varepsilon_{geometric} = \{\varepsilon_{isolated}, \varepsilon_{associated}\} \qquad (1)$$

$$\varepsilon_{isolated} = \{\varepsilon_{flat}, \varepsilon_{round}, \varepsilon_{\text{cylindricity}}, \varepsilon_{line}, \varepsilon_{planar}\} \qquad (2)$$

$$\varepsilon_{associated} = \{\varepsilon_{para}, \varepsilon_{perpen}, \varepsilon_{\text{inc}}, \varepsilon_{coa}, \varepsilon_{sym}, \varepsilon_{\text{loc}}, \varepsilon_{\text{cir}}, \varepsilon_{\text{total}}\} \qquad (3)$$

For each layer of 3D manifold model, the deformation $\varepsilon_{geometric}$ can be succinctly defined.

$$\varepsilon_{geometric} = max(\varepsilon_{A_i}) = max(\|(\varepsilon_{A_{i,x}}, \varepsilon_{A_{i,y}})\|) \qquad (4)$$

where $\|\ \|$ is 2-norm, $\varepsilon_{A_{i,x}}$ is the geometric error of $i$-th facet $A_i$ in $x$ direction and $\varepsilon_{A_{i,y}}$ is the geometric error of $i$-th facet $A_i$ in $y$ direction.

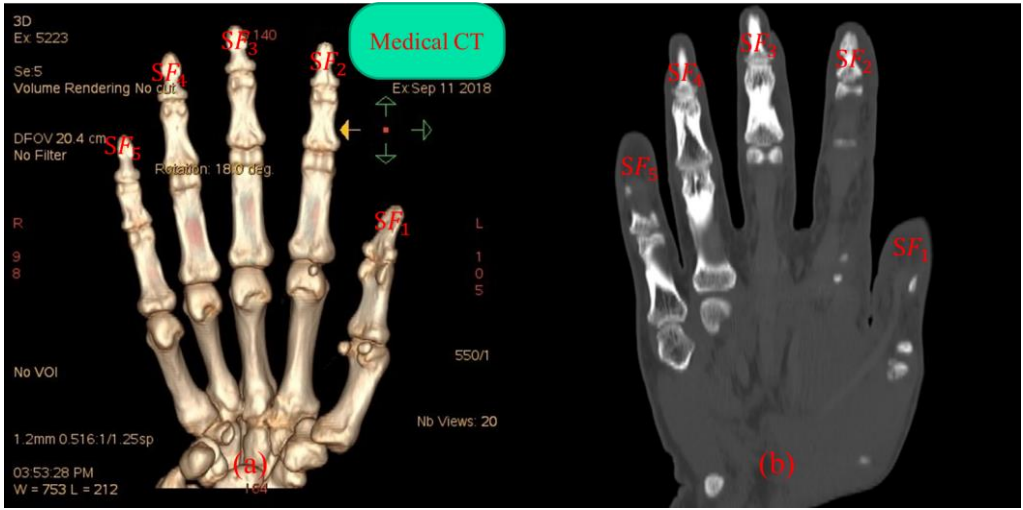

**Fig. 1** (a) Medical CT image of human hand;(b) Medical X-ray image of human hand.

Fig. 1 shows the Medical CT image and X-ray image of human hand. Flexible grasping space using oblique ellipsoids as convex hull. The skeleton lines of the five fingers can be thought of as ultimately meeting in a single point. Motivated by this, we seek to build a flexible grasping space based on these biomimetic skeleton lines. The mathematical approach we used to model 3DP process is firstly obtained by flexible grasping space (FGS) using oblique ellipsoids as convex hull. FGS allows to easily analyze multibody systems with pre-determined constraints. This is even more evident for the bionics structure, where the link's topologies and the joint's configurations do not change during the grasping process shown in Fig. 2. The specific steps of constructing oblique ellipsoids are as follows.

Given the oblique ellipsoid must be surrounded by a set of points P, and P is a subset of manifold vertices $V$, which means $P \in V$. The central equation of the ellipsoid is expressed as follows:

$$(x - P_s)^T \times A \times (x - P_s) = 1 \qquad (5)$$

where $P_s$ is the center of ellipsoid. In three-dimensional space, A is a $3 \times 3$ positive definite matrix. The eigenvectors of A define the principal axes of the ellipsoid and the eigenvalues of A are the reciprocals of the squares of the semi-axes.

The point set P is divided into a set of multiple sub points, and a bounding ellipsoid is generated for each subset. P is surrounded by a plurality of ellipsoid, and there will be some vertices of the facets are divided into different subsets. Through the detection of 3 vertices that cannot be in the triangle covered by a same ellipsoid, prevent these patches cannot be completely ellipsoid. Solving the minimum value of $\log det(A)$ to satisfy:

$$(P_g - P_s)^T \times A \times (P_g - P_s) \leq 1 \qquad (6)$$

where $P_g$ is the g-th column of the matrix P. $P_s$ is the center of the minimum volume enclosing ellipsoid (MVEE). T describes the matrix transpose. $det$ represents determinant of matrix. $A$ is the $3 \times 3$ matrix to be obtained.

The centroid of the triangle is calculated to find the ellipsoid with the minimum distance between the centroid and ellipsoid. According to the origin set of the selected ellipsoid and the 3 vertices of the triangle, the ellipsoid is recalculated. Multiple ellipsoids are obtained by iteration, until the error is less than the given envelope error. Finally, the MVEE is obtained, which covered the all facets completely.

Determining the parameters of ellipsoid equation A, $P_s$ . In the three-dimensional space, we can see that A is a $3 \times 3$ positive definite matrix by the definition of ellipsoid equation. Three matrices $U, Q, D$ are obtained when we complete singular value decomposition of the positive definite matrix A, where $U$, $D$ are Unitary Matrix respectively. The conjugate transpose of the Unitary Matrix is equal to its inverse matrix $U^H = U^{-1}$, $D^H = D^{-1}$, Q is the diagonal matrix of eigenvalues of A, $Q = diag(x_r^{-2}, y_r^{-2}, z_r^{-2})$. Finally, the parameters of ellipsoid equation are obtained. The standard equation of the ellipsoid is expressed as follows:

$$\frac{(x-P_s.x)^2}{x_r^2} + \frac{(y-P_s.y)^2}{y_r^2} + \frac{(z-P_s.z)^2}{z_r^2} = 1 \quad (7)$$

where $x_r$ represents the axis of the ellipsoid along the X direction, and $y_r$ represents the axis of the ellipsoid along the Y direction, and $z_r$ represents the axis of the ellipsoid along the Z direction.

Introducing the rotation angle of the three direction $\theta_x$, $\theta_y$, $\theta_z$. The oblique ellipsoid in the Cartesian coordinate system is constructed by means of affine transformation. $[x, y, z] \leftarrow [x, y, z] * R$. R describes the affine transformation matrix:

$$R = Rz \times Ry \times Rx \quad (8)$$

$$Rx = \begin{bmatrix} 1 & 0 & 0 & 0 \\ 0 & \cos(\theta_x) & -\sin(\theta_x) & 0 \\ 0 & \sin(\theta_x) & \cos(\theta_x) & 0 \\ 0 & 0 & 0 & 1 \end{bmatrix} \quad (9)$$

$$Ry = \begin{bmatrix} \cos(\theta_y) & 0 & \sin(\theta_y) & 0 \\ 0 & 1 & 0 & 0 \\ -\sin(\theta_y) & 0 & \cos(\theta_y) & 0 \\ 0 & 0 & 0 & 1 \end{bmatrix} \quad (10)$$

$$Rz = \begin{bmatrix} \cos(\theta_z) & -\sin(\theta_z) & 0 & 0 \\ \sin(\theta_z) & \cos(\theta_z) & 0 & 0 \\ 0 & 0 & 1 & 0 \\ 0 & 0 & 0 & 1 \end{bmatrix} \quad (11)$$

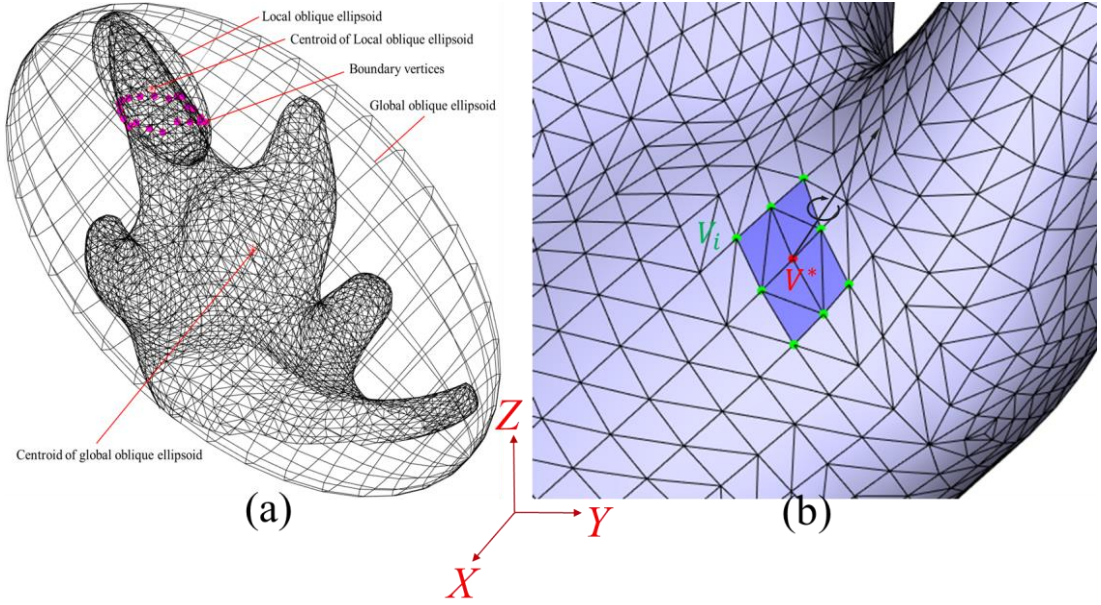

**Fig. 2** Schematic diagram of manifold morphing. (a) The global oblique ellipsoid and local oblique ellipsoid; (b) Red points denote control points $V^*$ and green points denote those points to be deformed.

## 2.2 Grasping space refining via manifold morphing

For a polyhedron manifold model $P_{object}$ in $R^3$, let $A_i$, $i = 0, \cdots, N-1$ be the N triangular faces of the $P_{object}$, with vertices $(a_i, b_i, c_i,)$, which are assumed to be ordered counter clockwise (CCW) on $A_i$.

The shape deformation regarding manifold morphing can be considered as a mapping from original to deformed space.

$$f: V_i \to V' \quad i = 1, \cdots, n \quad (12)$$

Instead of using absolute coordinates V, we would like to describe the manifold geometry using a set of differentials $\Delta = \{\delta_i\}$, Laplacian coordinate $\delta_i$ is used to describe the geometric manifold model, where $\delta_i$ is defined as follows:

$$\delta_i = L(V_i) = V_i - \sum_{j \in N_i} w_{i,j} V_j \quad (13)$$

where $L$ is the Laplacian operator of the manifold and $w_{i,j}$ is the weight of the vertex $V_j$ relative to the vertex $V_i$.

The combination of Laplacian - Gauss curve and uniform weight are selected as the

weight setting function:

$$w_{i,j} = e^{-j^2} * \frac{1}{card(N_i)} \qquad (14)$$

The degree $card(N_i)$ of this vertex is the number of elements in $N_i$. $e$ describes the natural base. Both the tensile and rotational morphing behaves better if the constraints are satisfied in a least squares sense, the morphing energy error equation is:

$$E(V') = \sum_{i=1}^{s} ||\delta_i - L(V')||^2 + \sum_{i=m}^{s} ||V' - u_i||^2 \qquad (15)$$

Among the morphing energy error equation, $V'$ describes the Euclidean coordinates of the deformed vertex $V_i$, $s$ describes the total number of vertices in the manifold model, $m$ describes the total number of vertices to be deformed, $V'$ describes a set of the euclidean coordinates of the deformed vertices. If $V_i$ is not to be deformed, $V'=u_i$ where $u_i$ describes the original coordinates of vertex $V_i$. If $V_i$ is to be deformed, first to distinguish and deal with the morphing types from control points $V^*$, and then proceed to the next step to solve for vertex $\{V'\}$, where $i \in \{1, ..., m-1\}$.

Differentiated processing and superimposing the vertices to be morphed and remaining vertices, the original manifold is decomposed into a low frequency base manifold (a manifold to be morphed) and a series of high frequency detail clusters (a non-morphing manifold), and the multi-resolution morphing of the model is obtained. Shaping energy under the corresponding stimulus plays a critical role in the maintenance of 3DP. On this ground, shaping energy or morphing energy, quantifies the disruption of intermolecular bonds that occur when a surface is created in 3DP process. In the physics of solids, surfaces must be intrinsically less energetically favorable than the bulk of a material (the molecules on the surface have more energy compared with the molecules in the bulk of the material), otherwise there would be a driving force for other surfaces to be created, removing the bulk of the material.

We solve the energy error functional $E(V')$ to get the minimum value. The vertices on the deformed manifold is forced to be close to the user specified position, but the other vertices are not changed. Finally, the wanted manifold model is obtained which is combined by low frequency manifold and high frequency detail cluster. The energy error functional $E(V')$ can be reverted to a sparse linear system:

$$\left(\frac{L}{0|I}\right)V' = \begin{pmatrix}\Delta\\U_i\end{pmatrix} \quad (16)$$

$L$ describes the coefficient matrix of Laplacian coordinate. $0$ describes the zero matrix. $I$ describes the unit matrix. $V'$ describes the Euclidean coordinates matrix of the deformed vertices. $\Delta$ describes the Laplacian coordinates matrix of the deformed vertices $U_i = \{u_i | i \in [m,\ s]\}$.

The method to solve the sparse linear system above is driving the LU (lower triangular and upper triangular) decomposition to the coefficient matrix L. Then the iterative method is used to obtain the Euclidean coordinates after the morphing of the deformed vertices $V'$. Finally, the all process of morphing is completed and shown as Fig. 3.

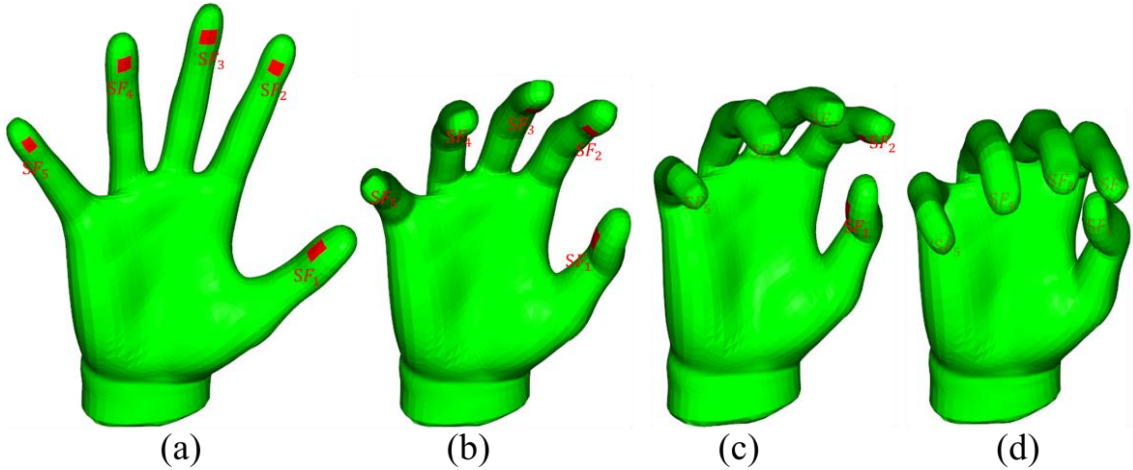

**Fig. 3** Original model and its self-grasping over time of right hand.

## 3  Kinematics model of self-grasping manifold

The structure of the self-grasping finger can also be abstracted into several linkages which are connected end to end. The geometric relationship between the linkages can be expressed by the position and posture relationship between the linkage coordinate systems fixed on each linkage mechanism shown as Fig. 4. Linkage transformation $^{i}T_{i+1}$ refers to spatial transformation relationship between linkage coordinate system $\{i+1\}$ relative to linkage coordinate system $\{i\}$ denoted as follows.

$$^iT_{i+1}(\theta_i) = \text{Rot}(z,\theta_i) \cdot \text{Trans}(z,d_i) \cdot \text{Trans}(x,a_i) \cdot \text{Rot}(x,\alpha_i)$$

$$= \begin{bmatrix} \cos\theta_i & -\sin\theta_i\cos\alpha_i & \sin\theta_i\sin\alpha_i & a_i\cos\theta_i \\ \sin\theta_i & \cos\theta_i\cos\alpha_i & -\cos\theta_i\sin\alpha_i & a_i\sin\theta_i \\ 0 & \sin\alpha_i & \cos\alpha_i & d_i \\ 0 & 0 & 0 & 1 \end{bmatrix} \quad (17)$$

where $\theta_i$ is joint rotation angle, $d_i$ is linkage offset, $\alpha_i$ is joint twist angle and $a_i$ is linkage length.

The forward kinematics model of the self-grasping finger is obtained by multiplying each linkage transformation in turn, and the end pose coordinate is obtained as follows.

$$\theta_t = {}^0T_t \cdot [\theta_0, \theta_1, \cdots, \theta_i]^T$$

$$= \left(\prod_{i=0}^{t-1} {}^iT_{i+1}(\theta_i)\right) \cdot [\theta_0, \theta_1, \cdots, \theta_i]^T \quad (18)$$

where $\theta_t$ is end pose coordinate, ${}^0T_t$ is end pose transformation matrix, and $t$ is the quantity of linkage coordinate system that means $\{0\}$ is finger base coordinate system and $\{t\}$ is fingertip coordinate system.

Differential kinematics of self-grasping finger is to study the relationship between the differential motion of the origin in fingertip coordinate system $\partial\theta_t$ that is relative to the fingertip coordinate system and the joint differential motion $\delta$. Let $\Theta = \{\theta_0, \theta_1, \cdots, \theta_i\}$ denotes one joint independent variable, the following conclusion can be obtained through the both sides derivation of forward kinematics model.

$$\partial\theta_t = \frac{\partial(\prod_{i=0}^{t-1} {}^iT_{i+1}(\theta_i))}{\partial\Theta} \cdot [\delta_0, \delta_1, \cdots, \delta_i]^T \quad (19)$$

In robotics, velocity Jacobian matrix is a transformation matrix that transforms the joint velocity vector into the generalized velocity vector that is relative to finger base coordinates system. Velocity Jacobian matrix $J_{f_i}$ of $f_i$-th self-grasping finger is defined from above equation as follows.

$$J_{f_i} = \frac{\partial(\prod_{i=0}^{t-1} {}^iT_{i+1}(\theta_i))}{\partial\Theta} \quad (20)$$

According to robotics theory, there is a dual relationship between differential kinematics and statics. It is assumed that the joints have no friction and the gravity of each member is ignored, then the static relationship can be established directly according to differential kinematics as follows.

$$\tau_{f_i} = J_{f_i}^T \cdot F_c^{f_i} \qquad (21)$$

where $\tau_{f_i}$ is generalized joint moment of the $f_i$-th self-grasping finger, $J_{f_i}^T$ is force Jacobian matrix that is the transposition matrix of velocity Jacobian matrix $J_{f_i}$ and $F_c^{f_i}$ is end force of the $f_i$-th self-grasping finger.

The operation control of self-grasping hand whose operation type is multi-finger hand not only needs to consider the movement and force of a single finger, but also must consider the synergism of movement and force between multiple fingers including statics, kinematics and dynamics.

Grasp statics studies the relationship between the contact force between finger and object and the force of object in multi finger grasping system. The contact between a finger and an object can be regarded as a mapping between the force exerted by the finger on the contact point and the resultant force of a reference point on the object. The contact models between fingers and objects include frictionless point contact, point contact with friction and soft finger contact. Self-grasping hand contacting with object is considered as soft finger contact.

In the soft finger contact mode, the finger can not only exert force in the friction cone, but also exert torque relative to the normal. In the case of multi-finger grasping, the mapping relationship of the contact force between fingers and objects is established through exerting the resultant force on the object by grasp matrix $G \in \mathbb{R}^{p \times m}$ as follows.

$$G: \mathbb{R}^m \to \mathbb{R}^p \qquad (22)$$

$$m = m_1 + m_2 + \cdots + m_k \qquad (23)$$

where $m$ is the dimension of contact force vector between all fingers and objects, $k$ is the number of fingers, $m_i(i = 1, \cdots, k)$ is the dimension of the contact force vector between the $i$-th finger and the object, and $p$ is degree of freedom of object.

$$F_o = G \cdot F_c \qquad (24)$$

where $F_o \in \mathbb{R}^p$ is resultant force on the object and $F_c = [F_c^1, F_c^2, \cdots, F_c^k]^T \in \mathbb{R}^m$ is contact force between fingers and objects.

The kinematics model of self-grasping hand is established as follows.

$$\begin{cases} G^T u = \dot{x} & (25) \\ J\dot{q} = \dot{x} & (26) \end{cases}$$

where $u \in \mathbb{R}^6$ is grasped object velocity vector, $\dot{x} = \left[\dot{x}_1^T, \dot{x}_2^T, \cdots, \dot{x}_k^T\right]^T$ is velocity vector of contact points, $\dot{x}_i \in \mathbb{R}^{m_i}$ is contact velocity vector in $i$-th contact point, $J = \text{diag}[S_1 J_1, S_2 J_2, \cdots, S_k J_k]$ is Jacobian matrix of self-grasping hand, $\dot{q}$ is velocity vector of joints, and $S_i$ is velocity selection matrix. In the soft finger contact mode, $S_i = \begin{bmatrix} I_{3\times 3} & 0 \\ 0 & 1 \end{bmatrix}$.

In Fig. 4(b), the centroid of flexible grasping space (FGS) is (0.0496, 0.1048, 0.1504), and its area and volume are 1.2484 and 0.0681 respectively. In Fig. 4 (b), the centroid (marked with grey) of flexible grasping space is (0.0724, 0.0006, -0.1660), and its area and volume are 0.8034 and 0.0495 respectively.

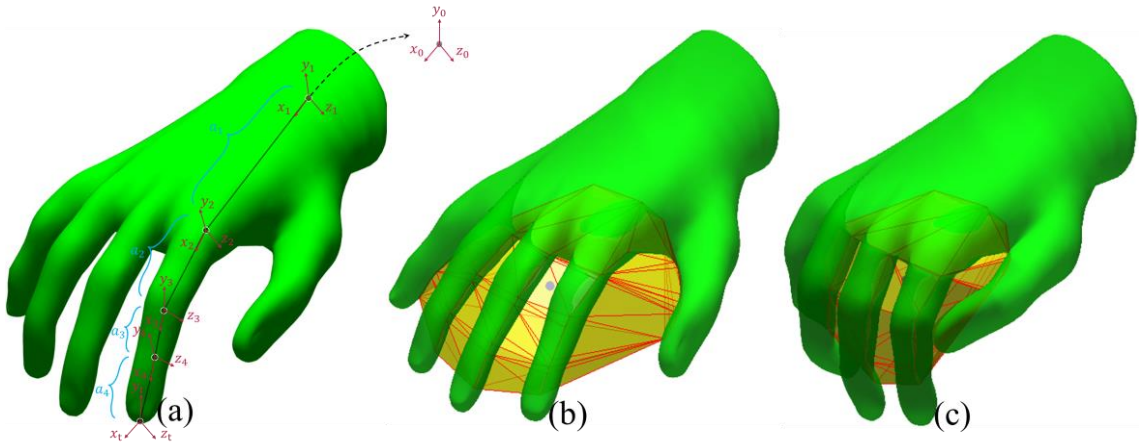

**Fig. 4** Diagram of Flexible grasping space (yellow) using oblique ellipsoids as convex hull (red). (a)Linkage coordinate system of self-grasping finger; (b) and (c) are two different grasping status of hand (green).

# 4 Energy management for 3D printing of manifold model

## 4.1 Fundament

Main components of total energy consumption $E_{total}$ for 3D printing come from melting energy consumption and mechanical movement. The melting energy consumption is as follows:

$$E_{melting} = cm_{total}(T_m - T_a) + m_{total}X = \rho V_T[c(T_m - T_a) + X] \quad (27)$$

where $E_{melting}$ is the melting energy consumption (kJ); $c$ is the material specific heat capacity $(J \cdot (kg \cdot K)^{-1})$; $m_{total}$ is the weight of the filament (kg); $T_m$ is the material melting point (K); $T_a$ is the environment temperature (K); $X$ is the latent heat $(kJ \cdot kg^{-1})$.

The total print time can also be estimated as follows:

$$t_T = \frac{L_T}{V_F} = \frac{r_{infill}V_T}{S_A V_F} \quad (28)$$

where $t_T$ is the total print time $(s)$; $L_T$ is the length of filament $(mm)$; $V_F$ is the printing velocity $(mm \cdot s^{-1})$; $r_{infill}$ is infill rate; $V_T$ is total volume of object to be printed; $S_A$ is cross area of filament.

In the actual measurement process, the working power for 3D printing is floating in real time. It is advisable that recording working power and print time in the meantime can calculate total energy consumption $E_{total}$.

$$E_{total} = \sum_{i=1}^{i=size(t_{sample})} \int_{t^{start}(i)}^{t^{end}(i)} P_{working}^{(i)} dt \quad (29)$$

where $t_{sample}$ is time sample sequence, $t_{sample} = \{t(1), t(2) \ldots t(end)\}$, and $P_{working}^{(i)}$ is working power sequence within the *i*-th sampling time.

For any 3D component to be fabricated, the object is sliced into a number of 2D layers of defined layer thickness. For the manifold model, the layered multi-polygons contain abundant geometric information which can be utilized to guide performance design. The Layered Customized Mask (LCM) of i-th layer can be obtained by converting multi-polygons $P_j$ into image set.

The Axis-aligned Bounding Boxes (AABB) of manifold model $P_{object}$ is generated to define the scale of the model. The stroke length $x_b, y_b, z_b$ of the AABB along $x, y, z$ direction can be calculated. The maximum print space for a printer along $x, y, z$ direction are respectively $x_p, y_p, z_p$. The oriented bounding box (OBB) means the bounding box with the minimum volume.

The normalized printing height ratio $h_n$ of the i-th slicing plane is proposed to denote accurate printing position and make it easy to compare which can be defined by layer height $z_i$ and total height $z_b$.

$$h_n = \frac{z_i}{z_b} \qquad h_n \in [0,1] \qquad (30)$$

The signed area $S_{section}$ of the 2D cross-section with $n$ non-self intersecting polygons is as follows:

$$S_{section} = \sum_{i=1}^{n} sign(P_i) * S_{P_i} \qquad (31)$$

where $S_{P_i}$ is the signed area of i-th polygons $P_i$, $sign$ is a signum function, $sign(P_i) = 1$, when $P_i$ is Counterclockwise (CCW); $sign(P_i) = -1$, when $P_i$ is clockwise (CW).

## 4.2 Build the relations between self-grasping deformation and LCM via semi-supervised deep residual network

In the traditional multi-layer feedforward neural network training, with the increase of network layers and training parameters, due to the disappearance of gradient, gradient dispersion and network degradation, the error increases in the late training period. The difference between the semi-supervised deep residual network [30] and the traditional multilayer feedforward neural network is that a residual block structure is used to superimpose on the original input, which alleviates the gradient disappearance problem caused by increasing the depth in the deep neural network, and enables the network to learn more data features. Therefore, based on the LCM of each layer shown in Fig. 8, the relations between self-grasping deformation and LCM via semi-supervised deep residual network is constructed.

$$H_i = f(H_{i-1}, W_i, b_i) = f(W_i \otimes H_{i-1} + b_i) \qquad (32)$$

where $H_i$ is output feature of the i-th layer, $W_i$ is weight of the i-th layer, $H_{i-1}$ is output feature of the (i-1)-th layer, $b_i$ is bias of the i-th layer, $f$ is activation function, $\otimes$ is convolution operation.

On the basis of Eq.(32), the residual block structure is added to the multilayer feedforward neural network to obtain the mapping relationship between the input and output of the monolayer residual network, as shown in Eq.(33).

$$H_i = f(f(H_{i-1}, W_i, b_i) + H_{i-1}) \qquad (33)$$

where $f$ is RELU activation function as follows.

$$f(x) = max(0, x) \qquad (34)$$

The mapping relationship between the input and output of the two-layer residual network can be further derived, as shown in Eq.(35).

$$\begin{aligned}H_{i+1} &= f(f(H_i, W_i, b_i) + H_i) \\ &= f(f(f(H_{i-1}, W_i, b_i) + H_{i-1}), W_i, b_i) + f(f(H_{i-1}, W_i, b_i) + H_{i-1}))\end{aligned} \qquad (35)$$

In order to predict the energy consumption of 3D printing according to the hierarchical customized graph, the mean square error loss $L_{MSE}$ is used as the loss function of the training semi-supervised deep residual network.

$$L_{MSE} = \frac{1}{2k} \sum_{j=1}^{k} (\hat{y}_j^k - y_j^k)^2 \qquad (36)$$

where $k$ is the number of output layer neurons, $\hat{y}_j^k$ is the expected value of the $j$-th neurons in the output layer, $y_j^k$ is the predicted value of the $j$-th neurons in the output layer.

The essence of training the semi-supervised deep residual network is to find the optimal set of parameters $\mathbf{x}^*$ to minimize the loss of mean square error $L_{MSE}$, and then continuously reduce the gap between the expected value and the predicted value, as shown in Eq.(37). The random gradient descent method is used in this paper.

$$\mathbf{x}^* = A\operatorname{rg\,min}(\frac{1}{2k}\sum_{j=1}^{k}(\hat{y}_j^k - y_j^k)^2)  \qquad (37)$$

The semi-supervised deep residual network takes each layer nozzle temperature, temperature gradient, printing velocity as the optimization parameters and solved MOO using Non-dominated Sorting Genetic Algorithm-II to calculate thermal deformation.

$$\text{Find: } \mathbf{x} = [\text{FGS}, T_n, \nabla T, V_F, d]$$

$$\text{Minimize: } f(\mathbf{x}) = \{E_{total}, \text{E}(V'), \varepsilon_{geometric},\} \qquad (38)$$

$$\text{Subject to:} \begin{cases} \text{FGS} \in \text{MVEE} \\ \min(d) \leq d \leq \max(d) \\ \min(V_F) \leq V_F \leq \max(V_F) \\ \min(T_n) \leq T_n \leq \max(T_n) \end{cases}$$

where $d$ is layer thickness (mm); $V_F$ is printing velocity (mm·s$^{-1}$); $T_n$ is nozzle temperature (K); $\nabla T$ is temperature gradient (K·mm$^{-1}$); $t_T$ is total print time (s); $E_{melting}$ is melting energy consumption (kJ); $\mathbf{x}$ is parameters of printing process; $f(\mathbf{x})$ is objective function of printing process.

## 5  Case study

### 5.1 Fabrication experiment and measurement

The specimens considered in the present work were fabricated on a 3D printer by using the commercial TPU as the mode material and the support material shown as Fig. 5. As shown in Fig. 6. The laser reconstruction equipment includes trilinear coordinates measuring instrument and rotary laser scanner. The scanning mode has turntable scanning and fixed scanning. The scanning precision is less than 0.05mm. The scan size can vary from 30mm×30mm×30mm to 1200mm×1200mm×1200mm.

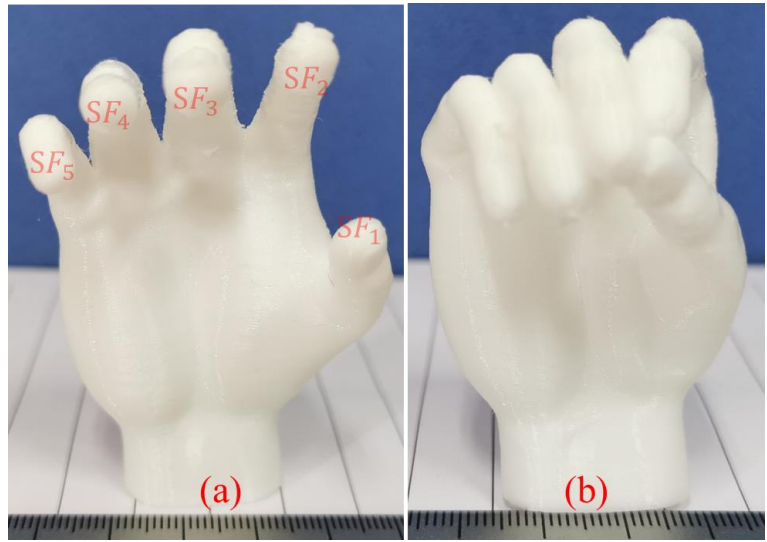

**Fig. 5** Physical functional hands in two different hand pose. (a) is claws pose and (b) is capisce pose.

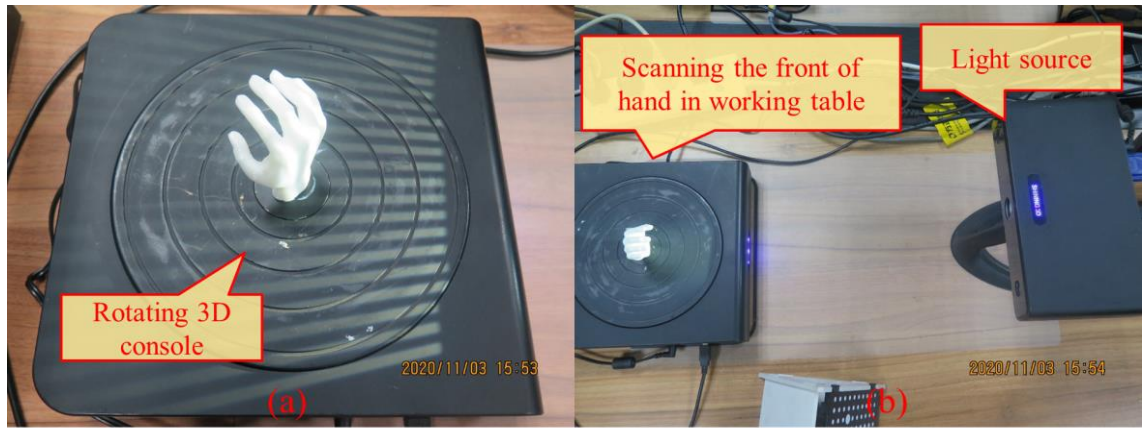

**Fig. 6** Rotary laser scanning of the self-grasping process. (a) Rotatable 3D console;(b) 3D laser scanner.

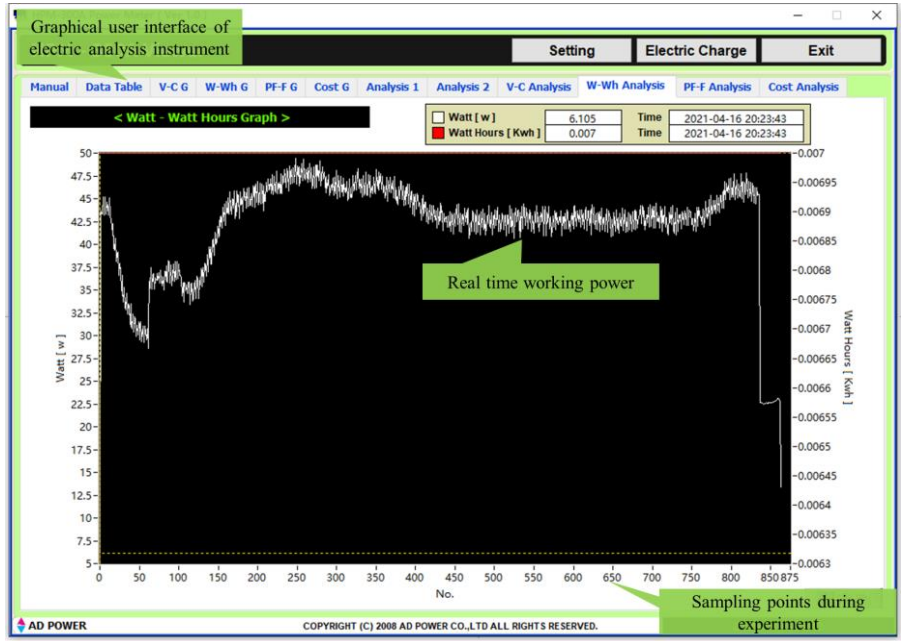

**Fig. 7** Graphical user interface of electric power analyzer.

The real time working power can be obtained at sampling points by electric power analyzer and graphical user interface (GUI) of electric power analyzer is shown as Fig. 7. Table 1 shows innovation comparison between the published papers and the proposed method.

### 5.2 Numerical results and comparison

The proposed method has been implemented using C++ and OpenGL. The proposed method was tested in a 64-bit Windows 7 system PC environment with Intel(R) Core(TM) i5-2310 CPU@2.9G, 2.9GHZ, NVIDIA(R) GTX 470 and 8G RAM.

The Axis-Aligned Bounding Box (AABB) in printing coordinate system (PCS) is obtained shown in Fig. 8, which bounding points are $V_1$=(29.2860,23.7940,27.9200), $V_2$ =(29.2860,23.7940,-42.4880), $V_3$ =(29.2860,-22.9780,27.9200), $V_4$ =(29.2860,-22.9780,-42.4880), $V_5$ =(-8.0880,23.7940,27.9200), $V_6$ =(-8.0880,23.7940,-42.4880), $V_7$ =(-8.0880,-22.9780,27.9200), $V_8$ =(-8.0880,-22.9780,-42.4880) respectively. The AABB center is (10.5990,0.4080,-7.2840) which ratio is (0.5308:0.6643:1). The printing

stroke space required for initial model is (37.3740,46.7720,70.4080) and the diagonal length of convex bounding box is 92.4214. The ratio of the center of gravity to the total length and height of the initial model is (71.8739%,51.2116%,48.2167%). Total surface area is 8033.7584, and total volume is 25352.4446. $L_T$ is total path length and $n_{point}$ is amount of turning points. Among all support structure length along z direction shown in Fig. 8(b), max is 65.0706, min is 0.0018, mean is 45.9816, median is 57.6240, sum is 143830.4183. Size(mass_center,1) is 3128, number of bottom support(whose bottom is marked with star *) is 294, number of non-bottom support(whose bottom is marked with square) is 2834.

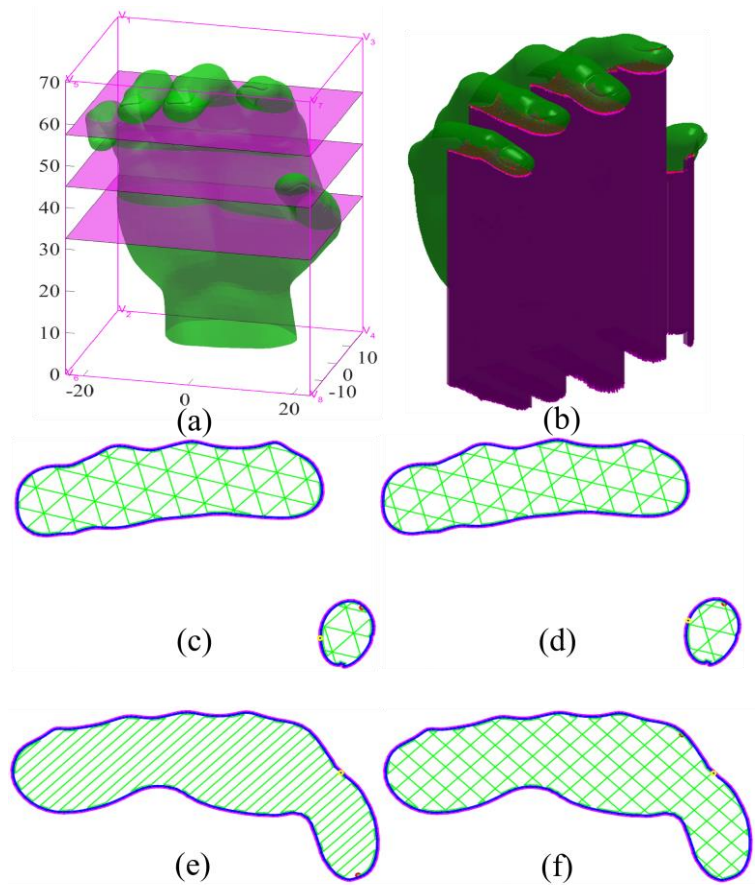

**Fig. 8** Layered slices obtained by converting multi-polygons into mask. (a) CAD model of hand with bounding box and slicing layers; (b) CAD model with support; (c) is triangle infill with $h_n = 0.64$, $L_T = 1440.919$ and $n_{point} = 45$; (d) is tri-hexagon infill with $h_n = 0.64$, $L_T = 1447.788$ and $n_{point} = 40$;. (e) is line infill with $h_n = 0.51$, $L_T = 1690.731$ and $n_{point} = 47$; (f) is grid infill with $h_n = 0.51$, $L_T = 1699.131$ and $n_{point} = 48$.

In order to prepare the 3D printing experiment and simulate the manufacturing process, the LCMs of specified location are shown in Fig. 9.

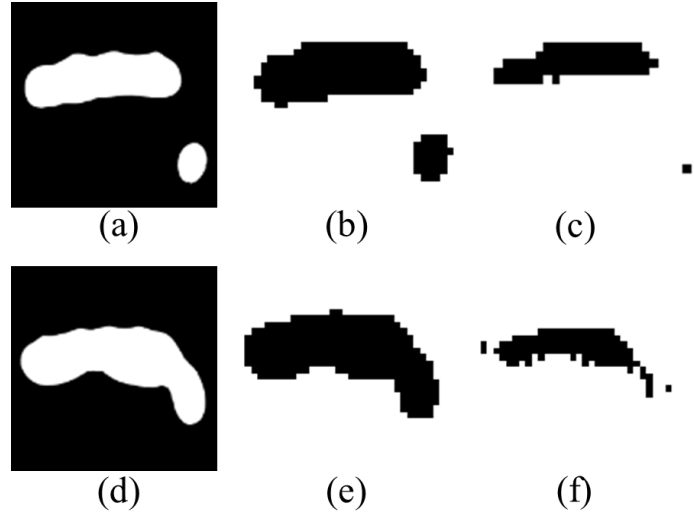

**Fig. 9** LCMs of specified location with $h_n$ 0.64 and 0.51 respectively where (a), (d) and (g) are original layered convolutional mask (LCM); (b), (e) and (h) are respectively LCMs with 32×32 resolution ratio; (c), (f) and (i) are multi-sclae convolution feature with dimension of 3×3 convolution kernal.

The numerical results of semi-supervised deep residual network for mechanical performance design are shown as Fig. 10. Within the uniform feasible range of training step, there exist minimum and maximum points which show aperiodic fluctuations without stairway. The training error overall declines, after several fluctuating downs and ups. Specifically, for theoretical value in Fig. 10(b), the global $\min(E_{total}) = 6$ at 100%; $\max(E_{total}) = 1172.50$ at 0.28%; $\text{mean}(E_{total}) = 367.56$. For predicted value in Fig. 10(b), the global $\min(E_{total}) = 216.56$ at 100%; $\max(E_{total}) = 460.89$ at 44.73%; $\text{mean}(E_{total}) = 355.52$. For theoretical value in Fig. 10(d), the global $\min(E_{total}) = 20$ at 100%; $\max(E_{total}) = 1172.50$ at 0.34%; $\text{mean}(E_{total}) = 423.86$. For predicted value in Fig. 10(d), the global $\min(E_{total}) = -33.65$ at 100%; $\max(E_{total}) = 675.81$ at 86.73%; $\text{mean}(E_{total}) = 417.53$.

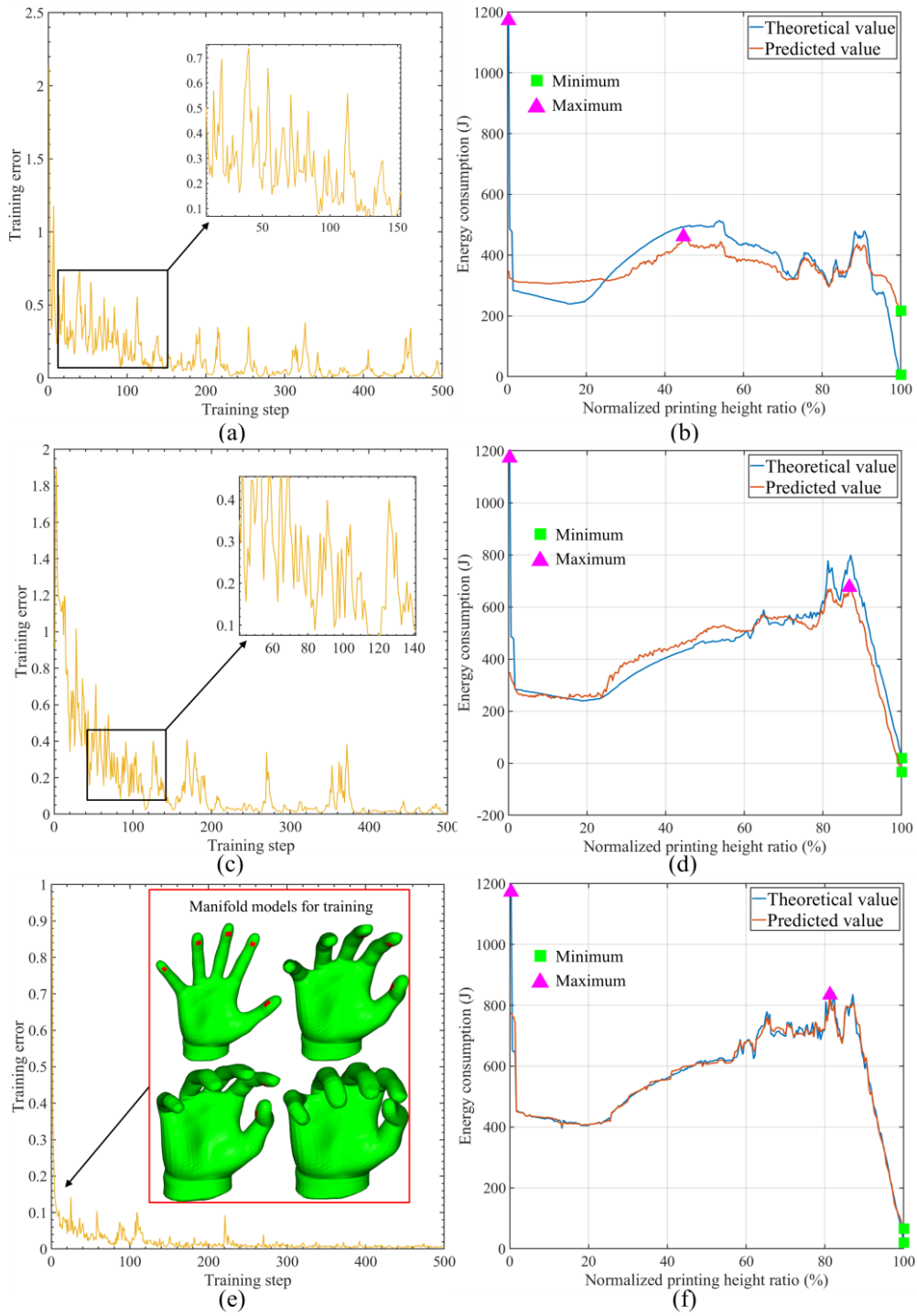

**Fig. 10** Numerical results of semi-supervised deep residual network for mechanical performance design. (a) and (b) is convergence process of loss function and comparison of theoretical and predicted values while using a model in Fig. 5(a) to train. (c) and (d) is convergence process of loss function and comparison of theoretical and predicted values while using a model in Fig. 5(b) to train. (e) and (f) is convergence process of loss function and comparison of theoretical and predicted values by using multiple manifold models which are generated by our method.

For theoretical value in Fig. 10(f), the global $\min(E_{total}) = 20$ at 100%; $\max(E_{total}) = 1172.50$ at 0.34%; $\text{mean}(E_{total}) = 558.58$. For predicted value in Fig. 10(f), the global $\min(E_{total}) = 65.90$ at 100%; $\max(E_{total}) = 834.23$ at 81.29%; $\text{mean}(E_{total}) = 558.06$. The result demonstrates that the semi-supervised deep residual network which receives multiple manifold models by our method performs better than other networks using a single model data to train. Table 1 further compares the innovations among the published papers and the proposed method.

**Table 1.** Innovation Comparison among the published papers and the proposed method.

| Concerns | The published papers | The proposed method |
|---|---|---|
| Grasps description | Feix et al. [31] synthesized human grasps into a single new taxonomy which considered only static and stable grasps performed by one hand. | ●Considering not only the various static and stable grasps, but also the dynamic grasps.<br>●It is suitable for different robotic grasping hands with multi fingers. |
| Flexible space description | Machine learning and particle swarm optimization to automatically pre-compute stable grasp configuration relying on GPU parallelization [32]. | ●No GPU parallelization need.<br>●Flexible grasping space (FGS) using oblique ellipsoids as convex hull shown in Fig. 3. |
| Grasp path planning | Grasp planning via hand-object geometric fitting and building a contact score map on a 3D object's | ●The proposed method can pre-compute grasp space before contacting. |

| | voxelization [33]. | ●Judging in advance whether the 3D object can be grasped to avoid redundancy calculation. |
| --- | --- | --- |
| Deep learning network | A large convolutional neural network to predict the probability that task-space motion of the gripper will result in successful grasps [34]. | ●The proposed semi-supervised deep residual network is more generative that relies on small samples. ●FGS provides prior knowledge before deep learning. |

# 6 Conclusions

Managing energy consumption in biological 3D printing system directly impacts the cost, efficiency, and sustainability of the process. In this paper, we propose an energy management system for 3D printing that leverages the refinement of manifold model morphing in a flexible grasping space. The manifold model is a mathematical representation of the 3D object to be printed, and the refinement process involves optimizing the morphing parameters of the manifold model to achieve desired printing outcomes. A flexible grasping space is shaped based on these biomimetic skeleton lines. We have obtained the further spatial expression of oblique ellipsoids convex hull. The proposed self-grasping method is able to deform the flexible hand from original pose to different grasping poses. The flexible grasping space of different grasping hands can be determined by using oblique ellipsoids as convex hull.

The semi-supervised deep residual network is employed to build the nonlinear implicit relations between self-grasping deformation and multi-scale features. The semi-supervised deep residual network takes each layer nozzle temperature, temperature gradient, printing velocity as the optimization parameters, and solved MOO using Non-dominated Sorting Genetic Algorithm-II to calculate self-grasping deformation. The proposed method is verified by a set of physical experiments. The trilinear coordinates measuring instrument and rotary laser scanner are used to determine the actual self-grasping deformation of the printed objects. Our method is more applicable for soft robotics and biomechanisms. Our proposed system addresses the challenges of limited sample data and complex morphologies of manifold models in layered additive manufacturing, resulting in less economical costs. The results highlight the importance of refining manifold model morphing in the flexible grasping space for achieving energy-efficient 3D printing, contributing to the advancement of green and sustainable manufacturing practices.

In future, the proposed method will be planned to be more suitable for 3DP of more complex part structures such as grooves, convex shoulders, matching holes and inner cavity microfluidic channels, supported by knowledge base in 3DP.

**Ethical Approval:** The type of study is non-human subject research, and waived the need for informed consent.

**Consent to Participate:** All authors have read and agreed to participate.

**Consent to Publish:** All authors have read and agreed to the published version of the manuscript.

**Competing Interests:** The authors declare no conflict of interests.
**Availability of data and materials:** Not applicable.

**Nomenclature**

$a^{Lr}(x)$——predicted output;

$b_i$——bias of feature map;

$b_j^k$——bias;

$c$——material specific heat capacity (kJ·kg$^{-1}$·K$^{-1}$);

$C_{Lr}$——error evaluation function;

$card(N_i)$ ——degree of this vertex;

$d$——layer thickness (mm);

$D$——discriminator;

e——natural base;

$E(V')$——energy error functional;

$E_{melting}$——hot-melt energy consumption (kJ);

$G$——generator;

$h_n$——normalized printing height ratio of the $i$-th slicing plane;

$H_i$——feature map of the $i$-th layer;

$L$——Laplacian operator;

$L_T$——total length of the filament (mm);

$m$——index of the first morphing vertex (kg);

$m_{total}$——weight of the filament (kg);

$\boldsymbol{n}$——unit normal vector;

$N_i$——set of adjacent vertices;

s——total number of vertices in the mesh model;

$sign$——signum function;

$S_{section}$——signed area of the 2D cross-section;

$S_{P_i}$——signed area of $i$-th polygons $P_i$;

$t_T$——total print time (s);

$T_a$——environment temperature (K);

$T_m$—— material melting point (K);

$T_n$——emperature of the nozzle (K);

$V$——mesh vertices set;

$V'$——euclidean coordinates set of the deformed vertices;

$V_F$——printing velocity (mm·s$^{-1}$);

$V_i$——the i-th mesh vertex;

$V_j$——the j-th mesh vertex;

$V_i'$——the i-th deformed vertex;

$V^*$——control point;

$x$——parameters of printing process;

$x_j^k$——the $j$-th neuron of the k-th layer;

$X$——latent heat (kJ·kg$^{-1}$);

$y(x)$——expected output in output layer $Lr$;

$z$——a noise vector from the distribution $p_z$;

$z_i$——layer height

$z_b$——total height

$\varepsilon_{geometric}$——geometric shape error which can be divided into isolated error $\varepsilon_{isolated}$ and associated error $\varepsilon_{associated}$;

$\varepsilon_{isolated}$——isolated error includes: flatness $\varepsilon_{flat}$, roundness $\varepsilon_{round}$, cylindricity $\varepsilon_{\text{cylindricity}}$, line profile $\varepsilon_{line}$, planar profile $\varepsilon_{planar}$;

$\varepsilon_{associated}$——associated error includes: parallelism $\varepsilon_{para}$, perpendicularity $\varepsilon_{perpen}$, inclination $\varepsilon_{\text{inc}}$, coaxiality $\varepsilon_{coa}$, symmetry $\varepsilon_{sym}$, location $\varepsilon_{\text{loc}}$, circular run-out $\varepsilon_{\text{cir}}$, total run-out $\varepsilon_{\text{total}}$;

$\varepsilon_{A_i,x}$——geometric error of $i$-th facet $A_i$ in $x$ direction;

$\varepsilon_{A_i,y}$——geometric error of $i$-th facet $A_i$ in $y$ direction;

$w_{i,j}$——weight of the vertex;

$w_{ji}^k$——weight assigned in the $j$-th neuron of the $k$-th layer to the input from $i$-th neuron of the previous layer;

$W_i$——weight assigned in the convolution kernel of the $i$-th layer;

$\delta_i$——Laplacian coordinate;

$\Delta$——set of differentials $U_i = \{u_i | i \in [m, s]\}$;

$\nabla T$——temperature gradient (K·$mm^{-1}$);

$\sigma$——hyperbolic tangent Tanh activation function;

$\rho$——material density (kg·$m^{-3}$);

$\sum L_i$——total servo trajectory

$\nabla_{\theta_d}$——gradient solving operator of discriminator $D$;

$\nabla_{\theta_g}$——gradient solving operator of generator $G$;